\documentclass[
superscriptaddress,
twocolumn,
showpacs,
bibnotes,
amsmath,
amssymb,
aps,
prm,
floatfix,
]{revtex4-2}
\usepackage{graphicx} 

\usepackage{graphicx}
\usepackage{dcolumn}
\usepackage{tabularx}
\usepackage{booktabs}
\usepackage{bm}
\usepackage{cleveref}
\usepackage{silence}
\usepackage{ulem}
\usepackage{color}
\usepackage{soul}
\usepackage{xcolor}
\usepackage{amsmath}

\newcommand{\yit}{Y(Nb$_{1-x}$Ta$_x$)$_6$Sn$_6$}
\newcommand{\lit}{Lu$_{1-y}$Y$_y$(Nb$_{1-x}$Ta$_x$)$_6$Sn$_6$}

\begin{document}

\title{Quantum oscillations and Dirac dispersion in tunable kagome lattice Lu$_{1-y}$Y$_y$(Nb$_{1-x}$Ta$_x$)$_6$Sn$_6$}

\author{Keenan Avers}
    \email[]{kavers@umd.edu}
    \affiliation{Maryland Quantum Materials Center and Department of Physics, University of Maryland, College Park, Maryland 20742, USA}

\author{Phineas Sobel}
    \affiliation{Maryland Quantum Materials Center and Department of Physics, University of Maryland, College Park, Maryland 20742, USA}

\author{Lochlan Joyce}
    \affiliation{Maryland Quantum Materials Center and Department of Physics, University of Maryland, College Park, Maryland 20742, USA}  

\author{Jared Dans}
    \affiliation{Maryland Quantum Materials Center and Department of Physics, University of Maryland, College Park, Maryland 20742, USA}      

\author{Prathum Saraf}
    \affiliation{Maryland Quantum Materials Center and Department of Physics, University of Maryland, College Park, Maryland 20742, USA}  

\author{Ram Kumar}
    \affiliation{Maryland Quantum Materials Center and Department of Physics, University of Maryland, College Park, Maryland 20742, USA}  

\author{Shanta Saha}
    \affiliation{Maryland Quantum Materials Center and Department of Physics, University of Maryland, College Park, Maryland 20742, USA}

\author{Peter Zavalij}
\affiliation{Department of Chemistry, University of Maryland, College Park, Maryland 20742, USA}

\author{Johnpierre Paglione}
    \affiliation{Maryland Quantum Materials Center and Department of Physics, University of Maryland, College Park, Maryland 20742, USA}
    \affiliation{Canadian Institute for Advanced Research, Toronto, Ontario M5G 1Z8, Canada}
    \email{paglione@umd.edu}
    
\date{\today}
    
\begin{abstract}
Kagome lattice crystal systems present interesting symmetry-protected band structure features such as flat bands, van Hove singularities, and linearly dispersing Dirac/Weyl points that provide a rich playground for strongly correlated electron physics.
Motivated by the rich properties and charge density wave evolution through the 1-6-6 series of compounds,
we present our results in single crystal growth and characterization of the of Lu$_{1-y}$Y$_y$(Nb$_{1-x}$Ta$_x$)$_6$Sn$_6$ double-alloy system, including evolution of the charge density wave transition, electrical transport behavior and resultant phase diagrams. 
Using a novel growth technique, the synthesis of high quality crystals with extended length along the crystallographic $c$-axis allows us to follow the gradual suppression of charge density wave (CDW) order with Y substitution, and observe quantum oscillations in both magnetoresistance and magnetization throughout the series.
We review the evolution of Fermi surfaces, effective masses and quasiparticle dispersion through the alloy series, revealing a decrease in size of Fermi surfaces that trends with both substitutions,  
and a scaling between effective mass and Fermi wavevector that suggests a regime with Dirac-like dispersion. 
The ability to fine-tune crystallographic, ground state and electronic dispersion properties of the \lit\ system with minimal impact of disorder opens a path torward further understanding the nature of the kagome lattice and its novel states and interactions. 


\end{abstract}

\maketitle

 
\section{Introduction}

The band structure of a material is inexorably linked to its crystal structure, which sets the symmetries and topological rules of its electron wave functions \cite{Singh2023AdvMat, Yin2025Symm_MathHeavyTopologicalReview, Brouwer2023Homotopic}. This enforcement ranges from the 3D heavy effective mass systems \cite{coleman2007heavyfermionselectronsedge, Yoshida2018NonHermitHeavyFermion}, to the competing orders and phases in cuprate superconductors \cite{Sebastian2015CuprateQuantumOsc}, and all the way down to effectively massless electrons of 2D graphene \cite{Castro2009RevModPhys.81.109GRaphene}. The kagome crystal lattice systems are built out of a motiff of corner sharing planner triangular atomic arrangements and enable particularly interesting band structure and topological features to manifest \cite{Wang2024AccMatRes_KagomeReviewFelser, Xia2025PRL_PhotonicKagome, Wang2023NatRevPhy_Kagome}. For non-magnetic kagome systems, the AV$_3$Sb$_5$ family has provided new avenues of insight into the interplay between band topology, charge density wave order (CDW), and unconventional superconductivity \cite{Zang2024PNAS_Cs135, Wilson2024NatRevMat135, Xu_2025, Zhang2023NanoLett_Cs135Flakes, Tan2023npjQuantMat_Cs135BerryCalc}. Magnetic kagome lattices bring in time-reversal-symmetry-breaking and enable tuning the band structure and Berry curvature in order to produce exotic Chern insulating states in Fe-Sn systems \cite{Ye2018Nature_Fe3Sn2, Kang2020Nature_FeSn}, as well as unconventional CDW order in FeGe \cite{Teng2023NaturePhysics_FeGe, Bonetti2025FeGeCDW_theory}.




There exist a few subtly different kagome lattice scions \cite{Sante2025RevModPhy_KagomeReview, Jovanovic2022_AmChemSoc_KagomeRules, Sharma2026npjQuantMat_TopologicalMaterialsReview}, however the RM$_6$X$_6$ family of materials- with R = rare-earth, M = transition metal, and X = Ga, Si, Ge, Sn - entails the widest scope of electronic, magnetic and topological properties. The ``166'' family has examples of both rare earth/actinide magnetism, such as the multiple triangular lattice magnetic phases of UV$_6$Sn$_6$ \cite{Thomas2025Nature_UV6Sn6} and UNb$_6$Sn$_6$ \cite{Riedel2025PRM_UNb6Sn6} which is no surprise given the awe-inspiring range of ground states of uranium systems \cite{Qiangqiang2025Science_UTe2, PhysRevMaterials.5.054803, Xu2024PRB_UBiTe_fluff}, as well as transition metal magnetism, such as spiral spin structures in YMn$_6$Sn$_6$ \cite{Nirmal2020SciAdv_YMn6Sn6_MagneticSprialStuffAndHall, Jia2024NanoLett_YMn6Sn6_STM}. Interplay between Mn magnetism and lanthanide magnetism enables investigation of large anomalous Hall effects with topology in RMn$_6$Sn$_6$ (R = Gd - Tm, Lu) \cite{Ma2021_RMn6Sn6} and disorder in related high entropy alloy form \cite{Min2022CommunPhys_HighEntrop166}. Heavy Fermion behavior alongside itinerant antiferromagnetism has been observed in YbV$_6$Sn$_6$  \cite{Lou2025PRL_YbV6Sn6}. Non-magnetic YV$_6$Sn$_6$ shows clean semimetallic behavior and also exhibits topologically non-trivial quantum oscillations  \cite{Pokharel2021PRB_YV6Sn6_GdV6Sn6}. 
Until recently, charge order has only been observed in ScV$_6$Sn$_6$ \cite{Arachchige2022PRL_Sc166_firstReport,Li2026SciChiPhy_Sc166Nernst, Saykin2026PRB_Cs135_Sc166_Kerr} and no reports of superconductivity have appeared for any 166 species to date. 

\begin{figure}[]
    \centering
    \includegraphics[scale=0.19]{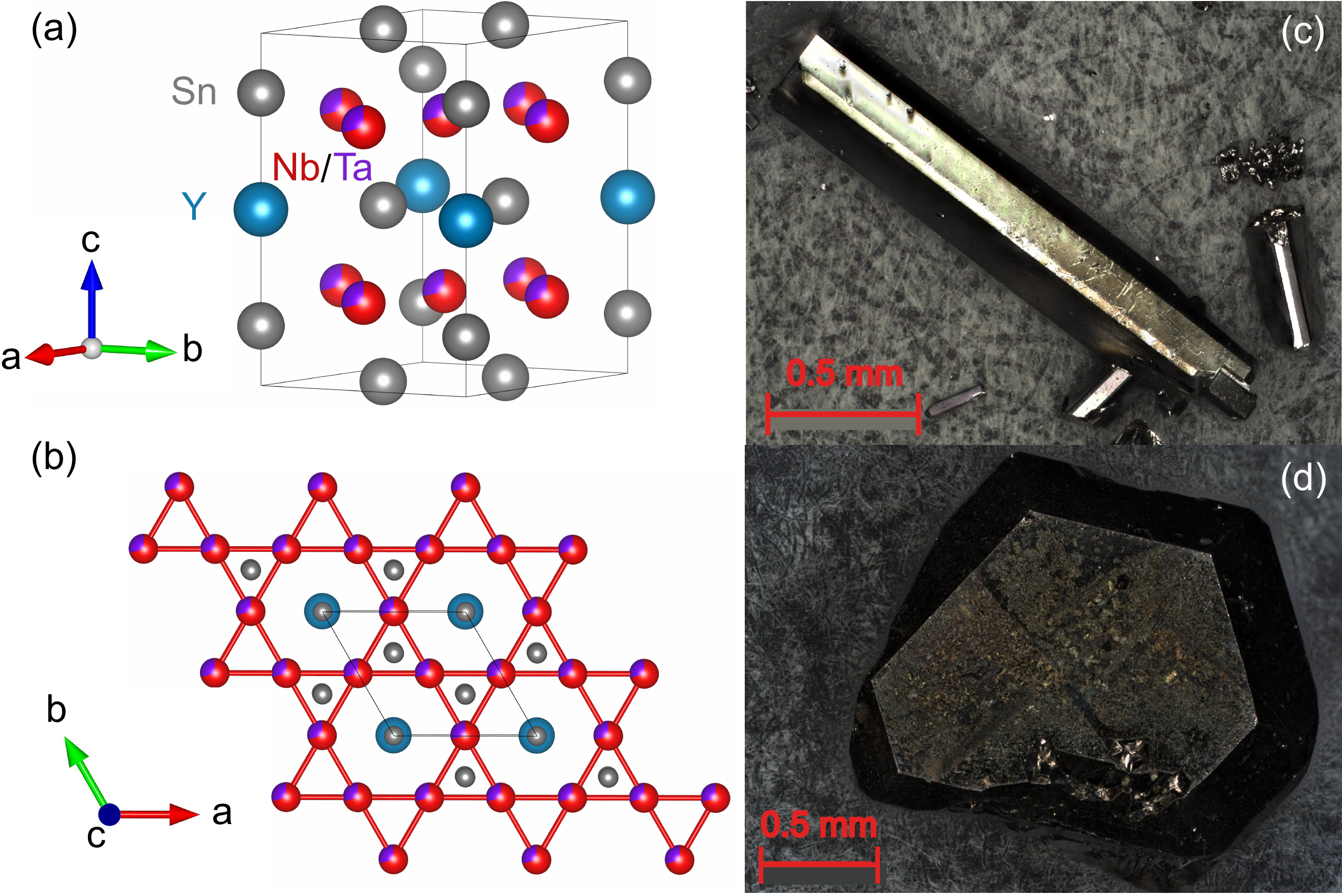}
    \caption{(a) The unit cell of \yit \hspace{1 pt} that shows the triangular Y layers sandwiched between kagome Nb/Ta  held together by a Sn framework. (b) Down $c$-axis view of extended crystal structure emphasizing the corner sharing motif of the kagome Nb/Ta sites. (c) Single crystal hexagonal needle of the 30 \% Ta batch. (d) Single crystal hexagonal plate of the 1 \% Ta batch that demonstrates Ta alloying affects growth morphology.}
    \label{fig:CrystalStruc}
\end{figure}

In this work we investigate the new Nb-based 166 family RNb$_6$Sn$_6$ \cite{Ortiz2025ACS_OriginalSynthesisLuNb6Sn6, XiaoPRB2025_GdNb6Sn6, Yue2012InorgChem_OriginalY166}, exploring a unique synthesis approach yielding a series of single-crystal specimens that incorporate both rare earth (Y,Lu) and transition metal (Nb,Ta) substitutions to form the system \lit \hspace{1 pt}.
The characterization of these crystals via electric transport are consistent with an anisotropic metal with easier electric current flow perpendicular to the hexagonal $c$-axis compared to along the $c$-axis. Despite this, we observe quantum oscillations in transverse magneto transport for the more resistive $c$-axis current that is tunable with minimal disorder from Ta-alloying. We also observe consistent quantum oscillations in magnetization across Lu-alloying. The oscillations indicate a system with cyclotron orbits of quasiparticles with $\sim 25\%$ bare electron masses and non-zero Berry phases. Upon alloying with Lu, the quantum oscillations exhibit increasing frequencies alongside abrupt changes as charge density wave (CDW) order emerges at the Lu end. A relation between effective mass and oscillation frequency establishes a non-parabolic band structure to one of the Fermi surfaces that is consistent with a linear Dirac dispersion at high carrier densities, but saturates to a constant effective mass as the  charge carriers are removed, which indicates a cross-over from relativistic Fermions to a more conventional massive particle almost certainly due to the presence of a gap in the Dirac band dispersion.

A single unit cell view of \yit \hspace{1 pt} is shown in Fig. \ref{fig:CrystalStruc} (a) that will be the focus of this work. Extended view along the $c$-axis in Fig. \ref{fig:CrystalStruc}(b) that emphasizes the kagome lattice motif of the Nb/Ta sites within the ab-plane. Photographs of hexagonal needle crystals from the 30 \% Ta batch in Fig. \ref{fig:CrystalStruc}(c) and hexagonal plates of the 1 \% Ta batch in Fig. \ref{fig:CrystalStruc}(d) demonstrate that Ta alloying has a subtle effect on the crystal morphology during growth. More importantly, we will show that this alloying with Ta also allows tuning the chemical potential and/or band structure near a gapped Dirac point with minimal disorder, which enables the observation of quantum oscillations in resistivity ($\rho$) and magnetization ($M$). We shall first focus on the \yit \hspace{1 pt} magneto transport and use that to set the stage for the \lit \hspace {1 pt} magnetization results in which we track the quantum oscillation evolution with both Ta and Lu alloying and eventual CDW onset at the Lu end.

\begin{table*}[ht] 
\caption{Single-crystal X-ray refinement parameters for \lit \hspace{1 pt} measured on a Bruker D8Venture w/ PhotonIII diffractometer at 295 K. Integral intensity were corrected for absorption using SADABS software \cite{Krause:aj5242} using multi-scan method. Structures were solved with the ShelXT \cite{Sheldrick:sc5086} program and refined with the ShelXL program \cite{Sheldrick:fa3356} using least-square minimization. All results are consistent with hexagonal space group P6/mmm with 1 formula unit per unit cell.}

\label{tab1}
\newcolumntype{C}{>{\centering\arraybackslash}X}
\begin{tabularx}{\textwidth}{CCCCCCC}

\toprule
x & 0.011(2) $\approx$ 0.01 & 0.295(4) $\approx$ 0.30 & 0.641(4) $\approx$ 0.64 & 0.004(2) $\approx$ 0.004 & 0.075(4) $\approx$ 0.08 & 0.021(2) $\approx$ 0.02 \\
y & 1.00 & 1.00 & 1.00 & 0.00 & 0.00 & 0.352(5) \\
\midrule
$a$ (\AA)     &   5.7583(1) & 5.7501(1) & 5.7422(1) & 5.7497(1) & 5.7451(1) & 5.7514(1)   \\
$c$ (\AA)     &  9.5195(3) & 9.5126(3) & 9.5122(3) & 9.5097(3) & 9.5110(3) & 9.5118(3) \\
$V$ (\AA$^3$) &  273.359(13)  & 272.384(13) & 271.596(13) & 272.263(13)   & 271.864(13) & 272.484(13)\\
$\rho$ (g/cm$^3$) &  8.288  & 9.233 & 10.377 & 8.824 & 9.066 & 8.687 \\
R$_1$ &  0.0100 & 0.0126 & 0.0113  & 0.0127 & 0.0151 & 0.0138\\
wR$_2$ &  0.0209 & 0.0278 & 0.0223 & 0.0239 & 0.0323 & 0.0277\\

\bottomrule
\end{tabularx}
\end{table*}

\section{Methods}

\subsection{Synthesis}


\begin{figure*}[]
    \centering
    \includegraphics[scale=0.33]{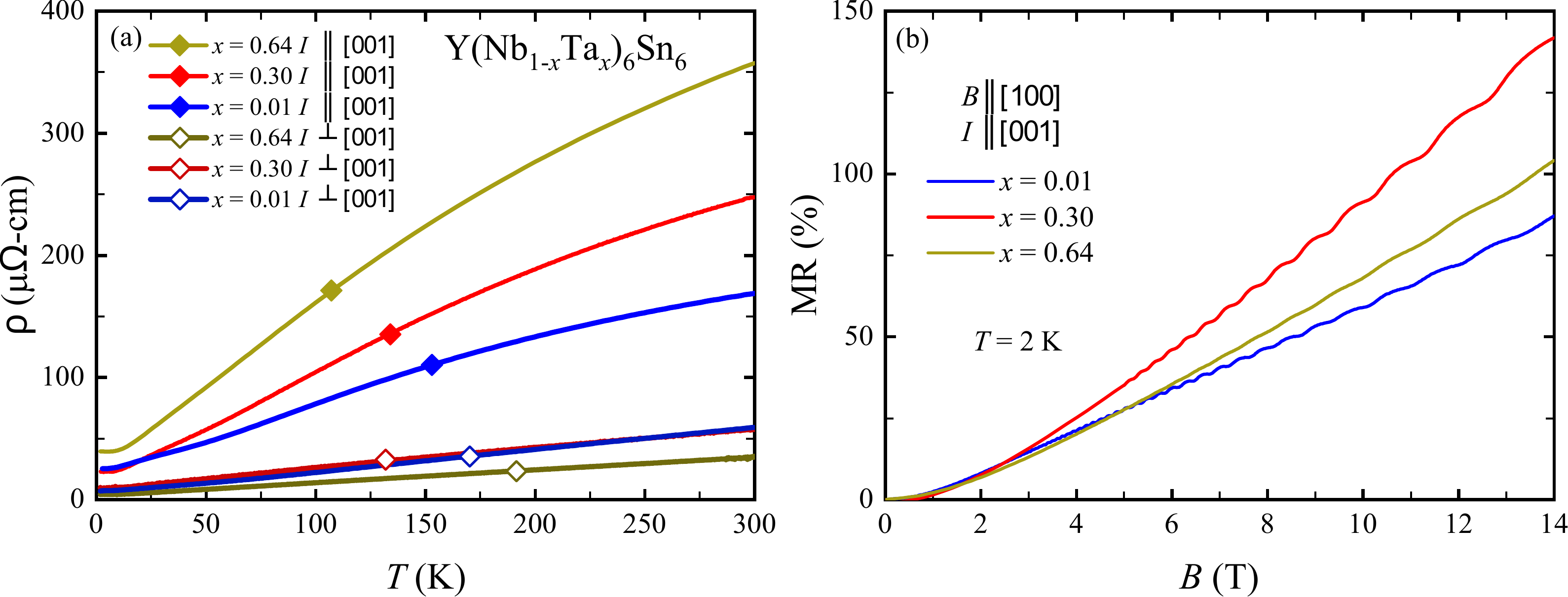}

    \caption{(a) Electrical resistivity ($\rho$) of \yit \hspace{1 pt} crystals  for current ($I$) along the $c$-axis ($I \parallel$ [001]) and perpendicular to the $c$-axis ($I \perp$ [001]) for the $x$ = 0.01, $x$ = 0.30, and $x$ = 0.64 batches.  The observation that \yit \hspace{1 pt} has smaller $\rho$ for $I \perp$ [001] compared to $I \parallel$ [001]  points to a reasonably large anisotropy. (b) The transverse magnetoresistance (MR) vs. magnetic field ($B$) for $B \parallel$ [100] and  $I \parallel$ [001] at $T$ = 2 K showing large MR response and low frequency Shubnikov-de Haas (SdH) quantum oscillations demonstrate high sample quality. The relatively moderate effect of Ta substitution on transport demonstrates minimal signatures of alloy disorder in both (a) and (b).}
    \label{fig:RvT_MR}
\end{figure*}

 Crystals of \yit \hspace{1 pt} were grown using an unconventional Sn flux technique. Initial amounts of Y, Nb and Sn were arc-melted into a button in an Ar atmosphere. This pre-reacted button was then flame sealed in an evacuated quartz tube with excessive amounts of Sn to act as a flux and Ta foil to encourage crystal nucleation. Most of the growths were performed in a single zone horizontal oven with the arc-melted button at the hot zone at 900 C$^\circ$. The crystals grew at the colder end of the ampule away from the hot zone, although as this is a single zone furnace we did not have good control nor characterization of the temperature gradient. Ta foil was placed at both the hot end and the cold end for the 30 \% ($x$ = 0.30) batch, but only at the cold end for the 1 \% ($x$ = 0.01) batch, which can be attributed for the difference in Ta \% alloying and morphological differences shown in Figs. \ref{fig:CrystalStruc}(c) and \ref{fig:CrystalStruc}(d). Attempts to grow 100 \% ($x$ = 1.00) have failed to result in any of the desired crystals and an attempt to grow explicitly targeted 95 \% ($x$ = 0.95) resulted in very small crystals of 64 \% ($x$ = 0.64) as a consequence of Ta having extremely minute solubility in molten Sn at these temperatures. It is an ongoing investigation as to how other variables such as temperatures, stoichiometry, masses, quartz tube length/geometry, etc influence the final crystal properties and quality. Samples appear air-stable and show no reaction to hot hydrochloric acid, which was used to etch away excess Sn. We observed that \lit \hspace{1 pt} analogs had similar behaviors and trends in terms of growth and handling, although we note that the hexagonal crystal plates appeared to become visibly thinner and smaller as Lu amount increased, as well as became more brittle and easier to break. We have also experimented with a two zone horizontal furnace for some of the \lit \hspace{1 pt} growths in which the hot zone was held at 900 C$^\circ$ and the cold end at 700 C$^\circ$ in which we observed the Ta amount could be below 1 \%, as checked by X-ray diffraction.

\subsection{X-Ray Diffraction}

 Single crystal X-ray diffraction was performed on crystals taken from a few of the batches and refined for Ta substitution on the Nb sites with the results shown in table \ref{tab1}. The actual final Ta percentages of the \yit \hspace {1 pt} are 1.1 \%, 29.5 \%, and 64.1 \% which we approximate as 1 \%, 30 \%, and 64\%, respectively,  for convenience and readability. A point worth mentioning is that the unit cell exhibits negligible volume change and may even shrink with more Ta substitution, despite Ta being the heavier 5$d$ analog of lighter 4$d$ Nb. This can be attributed to lanthanide contraction of the 4$f$ electrons bringing the atomic radii of Ta to be in line with that of 4$f$-free Nb \cite{Hand2022NbTaLanthanideContra_doi:10.1021/acs.inorgchem.2c02365}. For the purpose of this work, it now means we have a mechanism of tuning the Kagome band structure physics with minimal structural disorder owing to modulation in unit cell size throughout the crystal. The \lit \hspace{1 pt} crystals also exhibited a similar trend in unit cell size with an 8 \% Ta batch prepared in a single zone furnace, and a 0.4 \% batch prepared in a two zone furnace. It can be seen that the Lu end of the alloy has a smaller unit cell than the Y end. For the mixed Lu-Y batch, we note that although the final actual amount of Y was 35 \% ($y$ = 0.35), the initial nominal amounts of Y to Lu should have resulted in 25 \% ($y$ = 0.25). Such a stoichiometric mismatch can reasonably be attributed to different solubility limits of Y and Lu in molten Sn and differences in diffusion timescales between the two rare earth metals. For the rest of this manuscript we shall refer to samples by their $y$ and $x$ values with the $y$ value being determined from the initial nominal ratio of Y to Lu for said batch, and hence is an underestimate of the actual Y to Lu ratio based on the aforementioned $y$ = 0.25 batch. The $x$ amount shall be specified by the actual value extracted from X-ray refinement where applicable, although for batches in which a full refinement was not performed we shall simply use the notation of $x$ $<$ 0.1 for single zone furnace growths and $x$ $<$ 0.01 for the two zone furnace growths as good faith approximations consistent with the refined Ta amounts in Table \ref{tab1}.

\begin{figure*}[]
   \centering
    \includegraphics[scale=0.32]{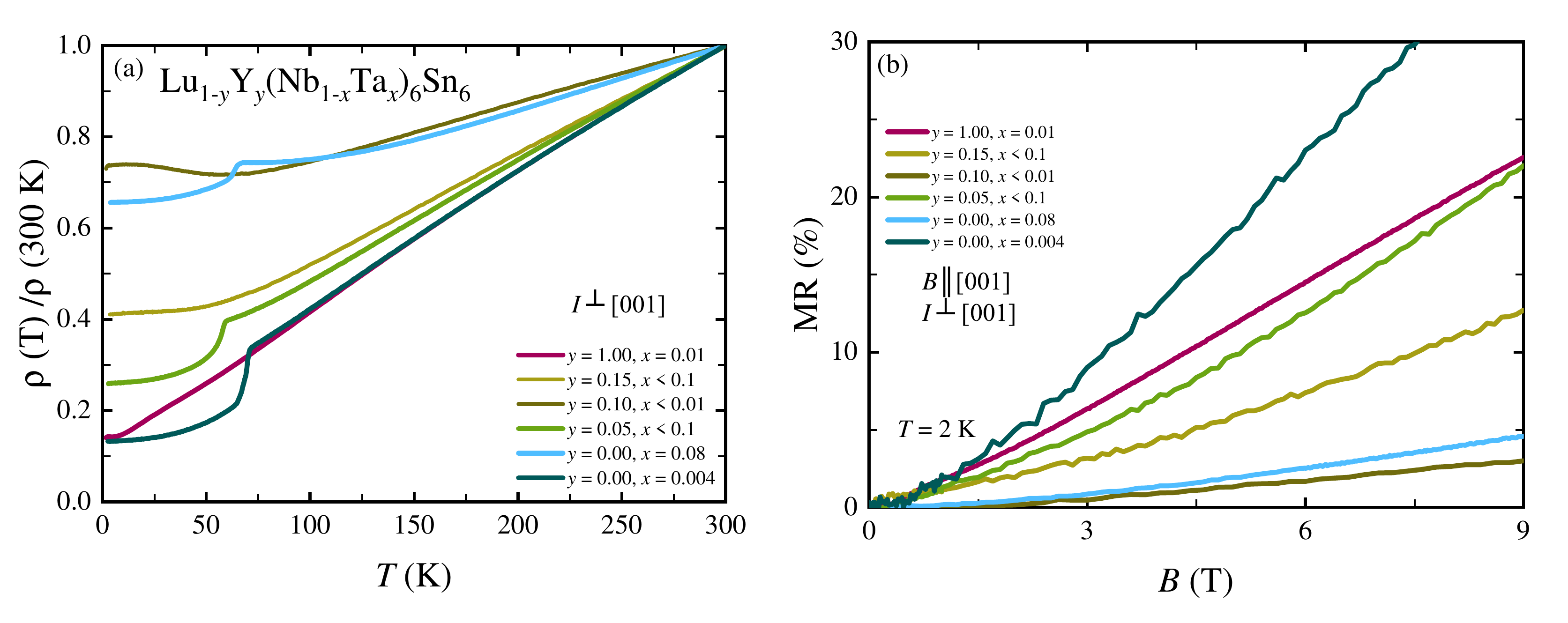}
    \caption{(a) Normalized resistivity of \lit \hspace{1 pt} for $I \perp$ [001]. The Y-rich end at $y$ = 1.00 exhibits good sample quality with minimal leveling off below $\sim$ 10 K, while samples with higher Lu concentration (e.g. $y\leq$0.15) show a consistent trend towards larger overall resistance ratio. Signatures of a charge density wave (CDW) emerge at $y$ = 0.10 and evolve from a local minimum towards a kink feature by $y$ = 0.00. (b) Magnetoresistance (MR) at $T$ = 2 K demonstrates a degree of consistency with Kohler's rule with the Y-rich samples having overall greater MR percentages, which suggests the increase in $\rho$ is due to disorder, although we emphasize that the $y$ = 0.00, $x$ = 0.004 batch demonstrates the largest MR response and nearly the cleanest sample among all batches investigated.}
    \label{fig:RhoAndMRFromYtoLu}
\end{figure*}

\subsection{Measurement Techniques}

 Electrical resistance ($R$) measurements were done on a commercial cryostat using the standard 4-point method with excitation currents ($I$) between 1 mA and 5 mA with no deviation from Ohmic behavior apparent. Artifacts from elemental Sn superconductivity have been manually deleted for aesthetics where applicable. Crystals of \lit \hspace{1 pt} were polished with conventional sand paper and alumina lapping films. Electrical contact was done with Ag epoxy (Epotek H20E) and 25 $\mu$m Ag wires and resulted in contact resistance of a few $\Omega$s. Magnetization ($M$) measurements were taken using the vibrating sample magnotometry technique up to 14 T on a commercial cryostat. A commercial SQUID magnetometer was also used up to 7 T.

\section{Results}

The electrical transport results of \yit \hspace{1 pt} vs. temperature ($T$) and magnetic field ($B$) are shown in Fig. \ref{fig:RvT_MR} and are consistent with a clean anisotropic metal with no strongly correlated thermodynamic order. In Fig. \ref{fig:RvT_MR} (a) we show the resistivity ($\rho$) vs. $T$ at $B$ = 0 for $x$ = 0.01, $x$ = 0.30, and $x$ = 0.64 with $I$ parallel to ($I \parallel$ [001]) and perpendicular ($I \perp $ [001]) to the $c$-axis ([001]) direction.
Among all the samples investigated, there is a consistent trend of transport perpendicular to the $c$-axis being more conductive, with $\rho$ (300 K) barely exceeding 50 $\mu \Omega$ - cm.  In contrast, electrical transport parallel to the $c$-axis is more resistive with $\rho$ (300 K) between 150 to 350 $\mu \Omega$ - cm. A few humps and inflection points between 75 K and 10 K for  $I \parallel$ [001] are of unknown origin, but show no signatures in other measurements and are likely due to relevant phonon modes being thermally depopulated and/or relative changes in mobility and densities of states between multiple bands. There is minimal indication of Ta alloying inducing significant disorder by changing the qualitative behavior of $\rho$, but may slightly raise residual resistivity at low $T$ from additional scattering.


 To this end, we present the $T$ = 2 K transverse magnetoresistance (MR) defined in the usual way as 100 \% *$(R(B) -R(0))/R(0)$ with $I \parallel$ [001] and $B \parallel$ [100] in Fig. \ref{fig:RvT_MR} (b). The MR of the $x$ = 0.01 sample is reasonably large, exceeding 75 \% by 14  T. If the MR magnitude is taken as a measure of sample quality, then it seems that the introduction of Ta has a inconsistent effect on mobility, as evidenced by the MR \% in the $x$ = 0.30 and $x$ = 0.64 sample being larger than the $x$ = 0.01 sample. This indicates that there are other factors that limit sample quality besides Ta substitution.  In all samples there are low frequency  Shubnikov-de Haas (SdH) quantum oscillations and is reminiscent of clean semi-metallic systems \cite{Vashist2019PRB_PrBi_QuantumOscillation, Singh_2020}. The power law behavior at large $B$ is clearly not the classic quadratic expectation, but rather is closer to linear \cite{Qiang2025PRB_L, Zhang2018_Bi2Se3_LinearMR}. Given the anisotropic nature of $\rho$ in Fig. \ref{fig:RvT_MR} (a), it also suggests that MR could be anisotropic as well.

 \begin{figure*}[]
    \centering
    \includegraphics[scale=0.33]{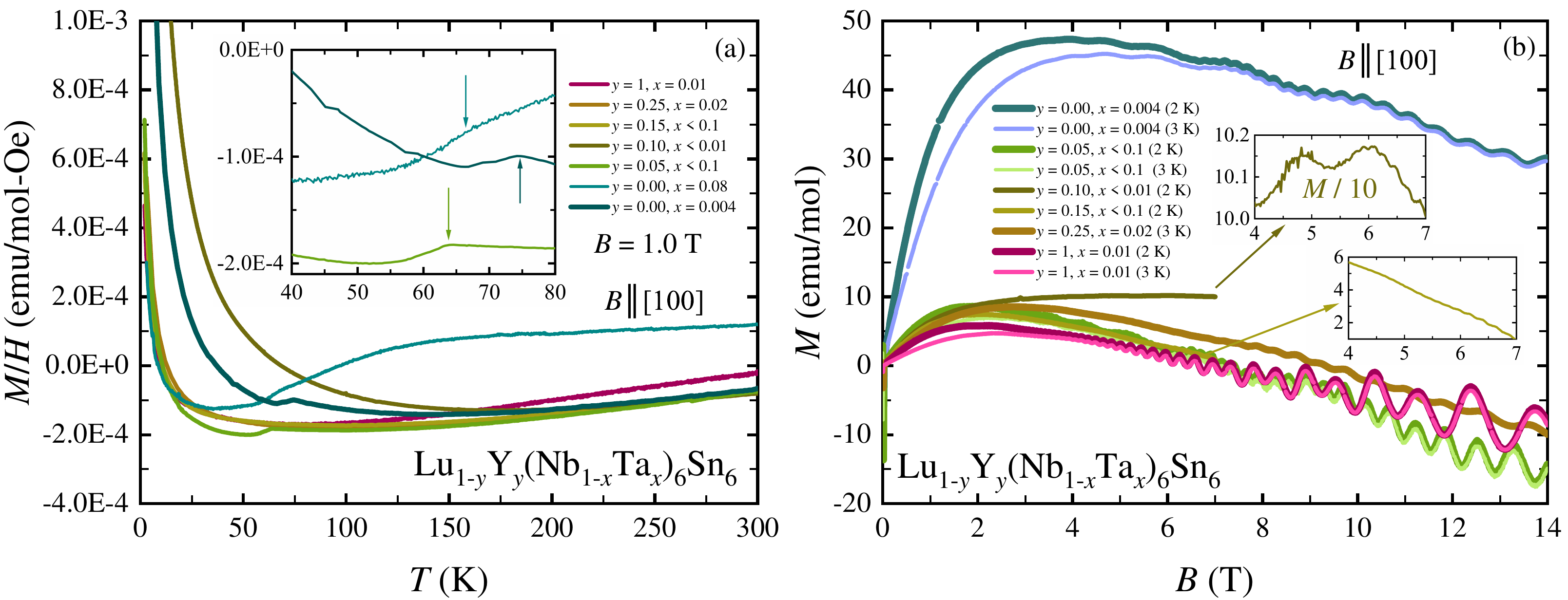}
    \caption{(a) Magnetic susceptibility ($M/H$) of various $y$ and $x$ batches of \lit \hspace{1 pt} single crystals measured in an applied field of 1.0 T applied along the [100] direction. The inset emphasizes the onset of the CDW inflection points in the Lu-rich batches. With the exception of the lower quality $y$ = 0.00, $x$ = 0.08 sample, all the materials are diamagnetic at 300 K, but become paramagnetic at low temperatures and fields. (b) Field dependence of magnetization ($M$) further demonstrates a mixture of paramagnetic and diamagnetic responses. The observation of de Hass-Van Alphen oscillations are consistent with quantum oscillations observed in resistivity (see text). The insets emphasize the low amplitude oscillations observed in the $y$ = 0.10 and $y$ = 0.15 batches, as well as the scaling of $y$ = 0.10 data down by a factor of 10 owing to a very large paramagnetic response.}
    \label{fig:VSM}
\end{figure*}

The $\rho  / \rho$ (300 K) vs. $T$ of \lit \hspace{1 pt} for $I \perp $ [001] is shown in Fig. \ref{fig:RhoAndMRFromYtoLu} (a) and shows a subtle evolution from $y$ = 1.00 to $y$ = 0.00. The Y-rich samples grow quite well and show minimal variation in sample quality. 
The $y$ = 0.15 sample starts a tendency of increased $\rho  / \rho$ (300 K) overall alongside a greater saturation in the low-$T$ limit. Signatures of a charge density wave (CDW) manifest at $y$ = 0.10 as a local minimum in $\rho$ and evolve towards a kink feature by the Lu end at $y$ = 0.00, which bear similarity to the behavior of LuNb$_6$Sn$_6$ under hydrostatic pressure \cite{Meier2025PRM_Lu166_PressureSurpressCDW}. We emphasize that the Lu-rich end does not necessarily need to be of poorer quality relative to the Y-rich end as evidence by the very high quality $y$ = 0.00, $x$ = 0.004 batch with $\rho  / \rho$ (300 K) magnitudes and trends comparable to the $y$ = 1.0 batches.

This sample quality variation across y and x are reflected  in the transverse MR for $I\perp$ [001] and $B \parallel$ [001] in Fig. \ref{fig:RhoAndMRFromYtoLu} (b) at $T$ = 2 K. Unlike the opposite $I$ and $B$ configuration for \yit \hspace{1 pt} in Fig. \ref{fig:RvT_MR} (b), the apparent behavior is closer to quadratic in $B$, of reduced overall \%, and suggests that the Fermi surfaces responsible for $I\perp$ [001] are topologically trivial.  Although an investigation of Kohler scaling \cite{Xu2021PhysRevX.11.041029KohlerScalingExtend} is not the focus of this work, a comparison of the the results between Fig. \ref{fig:RhoAndMRFromYtoLu} (a) and (b) at least suggests samples with lower \\ $\rho  / \rho$ (300 K) in the low-$T$ limit have larger MR response, especially the two $y$ = 0.00 samples with disordered $x$ = 0.08 having among the smallest response, while the clean $x$ = 0.004 has the largest \% within the given $B$ range. As conversion from $R$ to $\rho$ always has inherent uncertainties associated with how the geometry of voltage and current contacts are made \cite{Miccoli_2015_4pointAniversary}, especially with such small samples, we shall focus on the quantum oscillations moving forward.

The magnetization divided by induction field ($M/H$) vs $B$ of \lit \hspace{1 pt} is presented in Fig. \ref{fig:VSM} (a) for 1 T $\parallel$ [100]. Most of the samples are diamagnetic at 300 K and have a very small field response, consistent with a small density of states at the Fermi energy and a large inner core diamagnetic contribution from such heavy elements. The diamagnetic effect slightly increases upon lowering the temperature, perhaps as a consequence of losing contributions to Pauli susceptibility from thermally excited charge carries being depopulated across a small gap. The inset focuses on the  Lu-rich samples with kinks and inflection points in the data that indicate the onset of CDW order, although the signatures of such an order do seem to occur at slightly larger $T$ than the equivalent signatures in $\rho  / \rho$ (300 K) in Fig. \ref{fig:RhoAndMRFromYtoLu} (a). At the coldest temperatures measured there is a large upturn as the materials become paramagnetic. The trivial explanation for such an upturn is magnetic lanthanide impurities that accompanied the Y and Lu used in the growth.

 \begin{figure}[]
    \centering
    \includegraphics[scale=0.22]{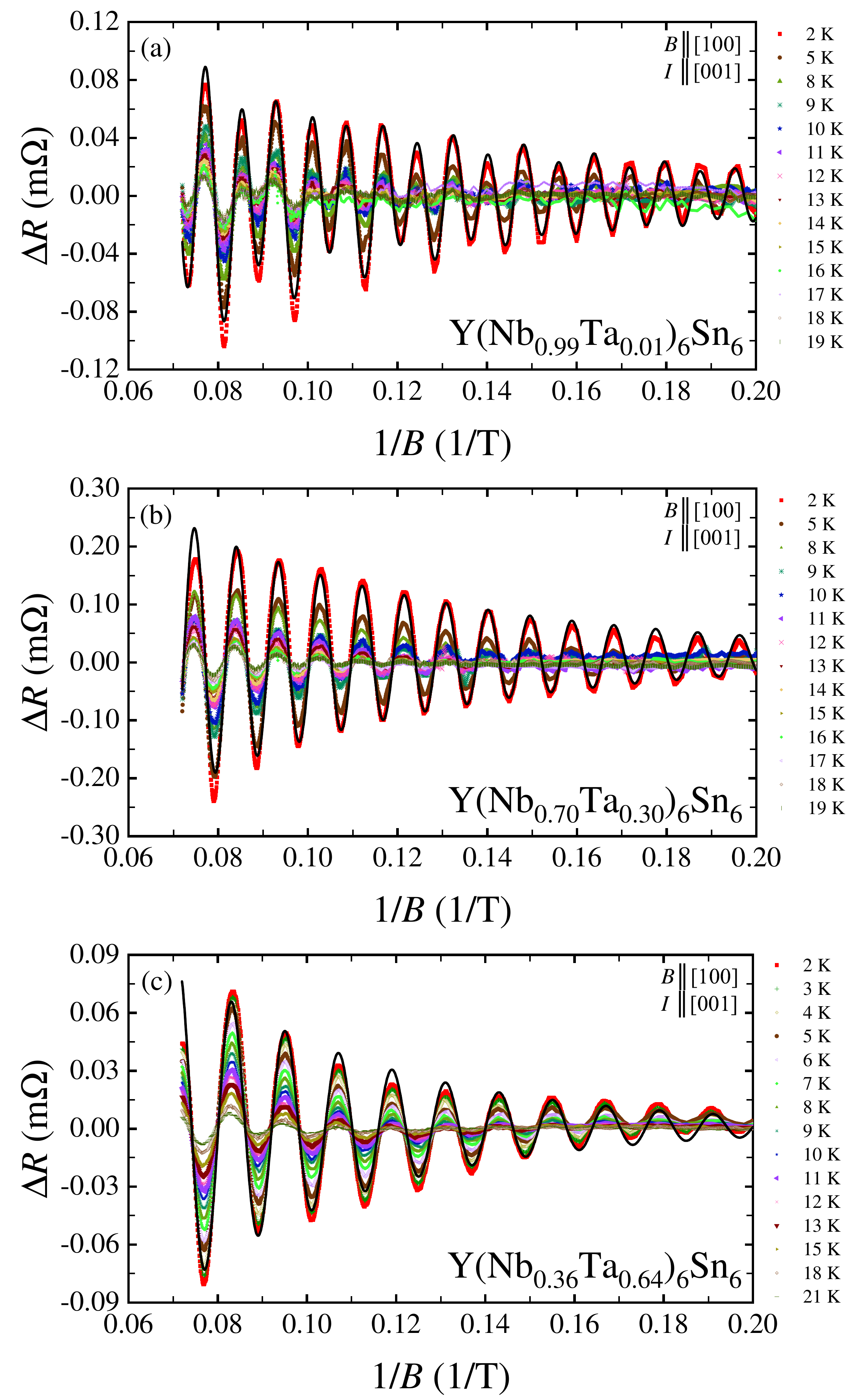}
    \caption{Shubnikov-de Haas (SdH) oscillations of three samples of 
    \yit, extracted by plotting background-subtracted magnetoresistance 
    $\Delta R$ (see text for details) vs inverse magnetic field, 
    for (a) $x$ = 0.01, (b) $x$ = 0.30, and (c) $x$ = 0.64. 
    Data fits to the Lifshitz–Kosevich formalism (see text and Table II) are shown for 2~K data (black curves). Two different frequencies, $\alpha$ and $\beta$, are resolvable for the $x$ = 0.01 sample, but only the $\alpha$ orbit is observable for the $x$ = 0.30 and $x$ = 0.64 samples. A non-zero Berry phase ($\phi_B$) manifests for all the orbits, which is consistent with a Fermi surface derived from topologically non-trivial linearly dispersing Dirac bands.
    }
    \label{fig:DRvInvB}
\end{figure}

The low-temperature $M$ vs. $B$ at in Fig. \ref{fig:VSM} (b) has  similar paramagnetic signatures at low field, although we note the contribution seems larger for some of the Lu-rich samples, which posits that there may be a non-trivial contribution to this paramagnetism related to a doping tuned density of states and sample quality and is not only from magnetic lanthanide contamination. The $y$ = 0.10 data in the upper inset had to be scaled down by a factor of 10 and has almost exact cancellation between the diamagnetic and paramagnetic contributions in the high field limit. With the exception of the $y$ = 0.00, $x$ = 0.08 (not shown), all samples are high enough quality to observe multiple-frequency de Hass-van Alphen (dHvA) quantum oscillations, although the reduced amplitude of the Lu-rich samples are consistent with increased defect density relative to the $y$ = 1.00 sample. The dHvA oscillations of the $y$ = 1.00, $x$ = 0.01 sample can be compared to the equivalent SdH ones in Fig. \ref{fig:RvT_MR} (b) and are of nearly the same frequencies, but out of phase with each other by $\pi$, as expected from the field-induced density of states modulation that is intrinsic to quantum oscillations, which shall now be analyzed and investigated further.

After fitting and subtracting polynomials to the $R$ vs. $B$ we obtain the change in $R$ ($\Delta R$) vs 1/$B$ in order to facilitate analysis of the SdH oscillations of \yit \hspace {1 pt} for the $x$ = 0.01, $x$ = 0.30, $x$ = 0.64 samples in Figs. \ref{fig:DRvInvB} (a), (b), and (c), respectively. We also interpolated the data to be evenly spaced in 1/$B$ to facilitate fast Fourier transform (FFT) analysis. There is a mix of two frequencies observable in the $x$ = 0.01 data, a greater amplitude big frequency on top of a lesser amplitude small frequency that we shall postulate correspond to the cyclotron orbits of the maximum and minimum orbits, respectively, of the same Fermi surface near a Dirac point. We shall name the big frequency orbit as $\alpha$ and the small frequency orbit as $\beta$. In contrast, for the $x$ = 0.30 data in Fig. \ref{fig:DRvInvB} (b) and the $x$ = 0.64 in Fig. \ref{fig:DRvInvB} (c) only single frequencies are resolvable, which suggests that the increased Ta alloying has shifted the bands such that the $\beta$ frequency is no longer resolvable. An important point to note is that these frequencies are on the order of $\sim$ 100 T, which are quite small Fermi surfaces, but ARPES and DFT results from LuNb$_6$Sn$_6$ show the presence of small Fermi surfaces at the Brillouin zone boundary K-point \cite{Yang2025PRB_Lu166_ARPES_Xray_FrustraitedCDW}. We shall show later that these frequencies continuously evolve from the Y system to the Lu system and hence these K-point small Fermi surfaces are strong candidate for the origin of these quantum oscillations.

In order to facilitate analysis we apply the Lifshitz–Kosevich (LK) equation used to analyze quantum oscillations of ScV$_6$Sn$_6$ from Ref. \cite{Zheng2024JPCM_Sc166_QO}. In total, the $\Delta R$ vs 1/$B$ can be fit to 
\begin{multline}
\Delta R = R' + \\
(B)^{1/2}R_{\alpha} R_{T\alpha} R_{D\alpha} *cos \left( 2\pi(\frac{F_{\alpha}}{B} - \sigma_{\alpha} +\delta_{\alpha} +\epsilon_{\alpha}) \right) \\
 +(B)^{1/2}R_{\beta} R_{T \beta} R_{D\beta} *cos \left( 2\pi(\frac{F_{\beta}}{B} - \sigma_{\beta} +\delta_{\beta} +\epsilon_{\beta})\right)
\label{eqn:FullLK}  
\end{multline}
in which $R'$ is a constant offset, $R_i$ ($i = \alpha,\beta$) are the amplitudes for the orbits and relate to the respective density of states, $R_{Ti}$ are the reduction factors due to finite temperature, $R_{Di}$ is the Dingle term that accounts for scattering, $F_i$ are the frequencies in T that correspond to the Fermi surface cross sectional area, $\sigma _i$ are the phase shifts that correspond to the Berry phases ($\phi_{Bi}$) via $\phi_{Bi} /2\pi = \sigma_i - 1/2 $, $\delta _i$ are the phase shifts from the dimensionality of the Fermi surface in question with $\delta_{\alpha} = -1/8$ being the contribution from a maximum orbit of a 3D surface while $\delta_{\beta} = +1/8$ for the minimum orbit of a 3D surface, and $\epsilon_i = +1/2$ under the assumption that $\rho_{xx} (B) >> \rho_{xy} (B)$ (ie the Hall contribution is negligible compared the resistive contribution), although we note that we lack measurements of the Hall effect in our samples in this current and field configuration. The Dingle term is given by
\begin{equation}
R_{Di}=exp(-\Gamma m_{i}T_{Di} /B)
\label{eqn:DingleOnly}  
\end{equation}
 with $\Gamma =2 \pi^2 k_B m_e/(e \hbar) \approx 14.69$ T/K, $m_i$ is the effective mass of the orbit relative to the bare electron mass ($m_e$),  and $T_{Di}$ is the Dingle temperature, proportional to the scattering strength. The temperature reduction factor term is given by 
\begin{equation}
R_{Ti}= \frac{\Gamma m_i T/B}{sinh(\Gamma m_i T/B)}
\label{eqn:TemperatureOnly}  
\end{equation}
and serves as a way to extract the effective masses. We emphasize that attempting to fit eqn. \ref{eqn:FullLK} to $\Delta R$ vs. 1/$B$ is not viable as the $m_i$'s have dependencies with the other fit coefficients that make the usual least squares fit approach ill suited.

\begin{table*}[ht] 
\caption{Result of Lifshitz-Kosevish (LK) treatment to the $\Delta R$ vs. 1/$B$ data in Figs. \ref{fig:DRvInvB} and  \ref{fig:DR_FFT_EffectiveMass} as described in the main text.}

\label{tab2}
\newcolumntype{C}{>{\centering\arraybackslash}X}
\begin{tabularx}{\textwidth}{CCCCCCC}

\toprule
x & orbit & $m_{i}$ ($m_e$) & $F_{i}$ (T) & $T_{D i}$ (K) & $R_{i}$ (m$\Omega$/$\sqrt{\mathrm{T}}$) & $\phi_{B i} / 2 \pi$   \\

\midrule
0.01 &   i = $\alpha$ & 0.249(4) & 127.04(3) & 1.73(5) & 0.0358(7) & -0.304(3)  \\
0.01 &   i = $\beta$ & 0.152(3) & 67.7(2) & 6.3(4) & 0.018(2) & 0.22(2)   \\
0.30 &  i = $\alpha$ & 0.237(1) & 106.93(3) & 2.99(7) & 0.126(3) & -0.135(4) \\
0.64 &  i = $\alpha$  & 0.257(3) & 83.36(5) & 4.00(8) & 0.069(2) & -0.194(4)\\

\bottomrule
\end{tabularx}
\end{table*}

\begin{figure}[]
    \centering
    \includegraphics[scale=0.23]{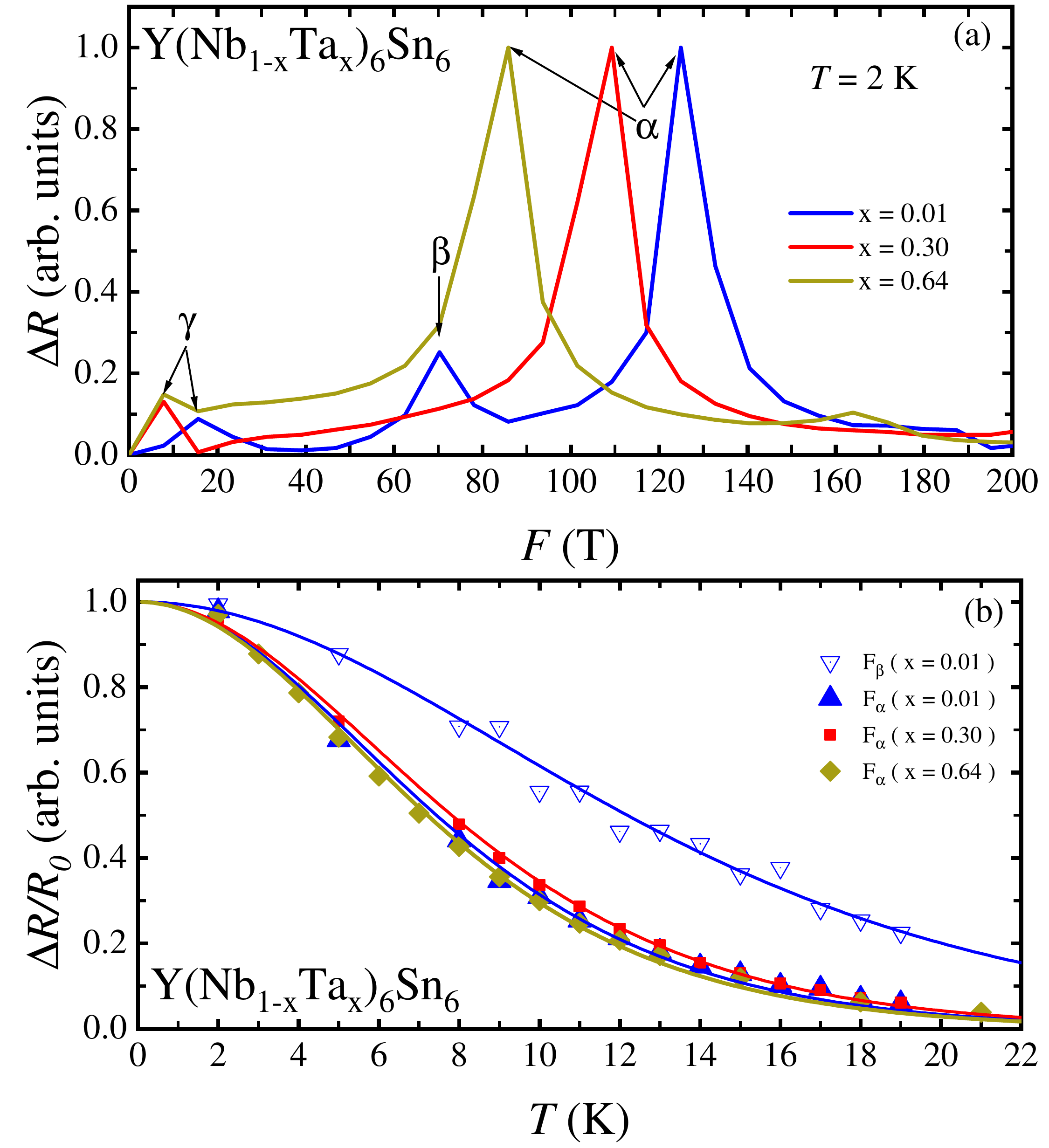}
    \caption{(a) Normalized Fast Fourier transforms of oscillatory component of magnetoresistance data of \yit \hspace{1 pt} from Fig. \ref{fig:DRvInvB}. The $\alpha$ and $\beta$ frequencies are labeled, but we also point out a low frequency peak $\gamma$ (see text for discussion). (b) Temperature dependence of the FFT peak heights normalized to zero field resistance ($R_0$). The solid curves are fits to the data used to extract the effective masses ($m$) as described in the text. 
    }
    \label{fig:DR_FFT_EffectiveMass}
\end{figure}

Instead, we took the fast Fourier transform (FFT) of $\Delta R$ vs. 1/$B$ between 14 T (0.0714 T$^{-1}$) and 5 T (0.2 T$^{-1}$), with a normalized $T$ = 2 K example shown in Fig. \ref{fig:DR_FFT_EffectiveMass} (a), and extract the amplitude of the relevant frequency components at each constant temperature data set in Figs. \ref{fig:DRvInvB} (a), (b), and (c).  The frequency ($F$) components are labeled, but we note the existence of a feature labeled $\gamma$ present at low frequency that can easily be discarded as an artifact of background subtraction, but we shall show  later that this is actually a real orbit that evolves with Lu alloying. The frequency resolved amplitudes can then be normalized by $R$ at $B$ = 0  ($R_0$) and fit as a function of $T$ via eqn. \ref{eqn:TemperatureOnly} to extract $m_i$ \cite{Murakawa2013BerryPhaseRashbaSemiconductor}. The results of this approach are shown in Fig. \ref{fig:DR_FFT_EffectiveMass} (b) and are retroactively normalized to 1 at 0 K for clarity with the effective masses in table \ref{tab2}. We observe that the $x$ = 0.01 sample has two different effective masses with the smaller $\beta$ frequency component having a lighter effective mass of 0.152(3) $m_e$ compared to the larger $\alpha$ frequency with a heavier effective mass of 0.249(4) $m_e$. As only a single frequency is observable for the $x$ = 0.30 and $x$ = 0.64 samples we can only obtain the $\alpha$ frequency orbit effective masses of 0.237(1) $m_e$ and 0.257(3) $m_e$, respectively. 

We take these fixed effective masses and plug them into Eqn. \ref{eqn:FullLK} and then fit them to the $\Delta R$ vs. $1/B$ at 2 K in Figs. \ref{fig:DRvInvB} (a), (b), and (c) with $R'$, $R_i$, $T_{Di}$, $F_i$, and $\sigma_i$ as free fit parameters. With appropriate initial guesses, a standard least-squares fitting process could produce reasonable results as shown in the black curves with the fit coefficients in table \ref{tab2}. The results seem to suggest that the $\beta$ frequency component has both a smaller amplitude ($R_{\beta}<R_{\alpha}$) and more scattering ($T_{D \beta}> T_{D \alpha}$), which may explain why we are unable to observe the small oscillations in the $x$ = 0.30 and $x$ = 0.64 samples; both an overall smaller density of states and increased scattering could mask it. An important takeaway is that $\phi_{Bi}$ takes on non-zero values, which indicates non-trivial topology could be present in the bands that the SdH oscillations are observed in. Although the Berry phase should be the topologically protected $\pm \pi$ for a linear dispersing band \cite{Alexandra_2018_PRB_TopoMetalTheory_PhysRevB.97.144422}, there are additional contributions to the shift that can come from deviations from linearity \cite{Taskin2011PRB_BerryNonIdeal}, Zeeman splitting \cite{Wang2023_PRM_RhSb3, KOSSUT1978SolidStateComm_SpinSPlitQuantumOscillations, MartinPRB2013_RashbaSplitQuantumOscillations_BiTeI}, and strong spin-orbit coupling \cite{Tan2023npjQuantMat_Cs135BerryCalc}. A consistency check using a separate measurement technique will be helpful in this regard.

 \begin{figure}[]
    \centering
    \includegraphics[scale=0.29]{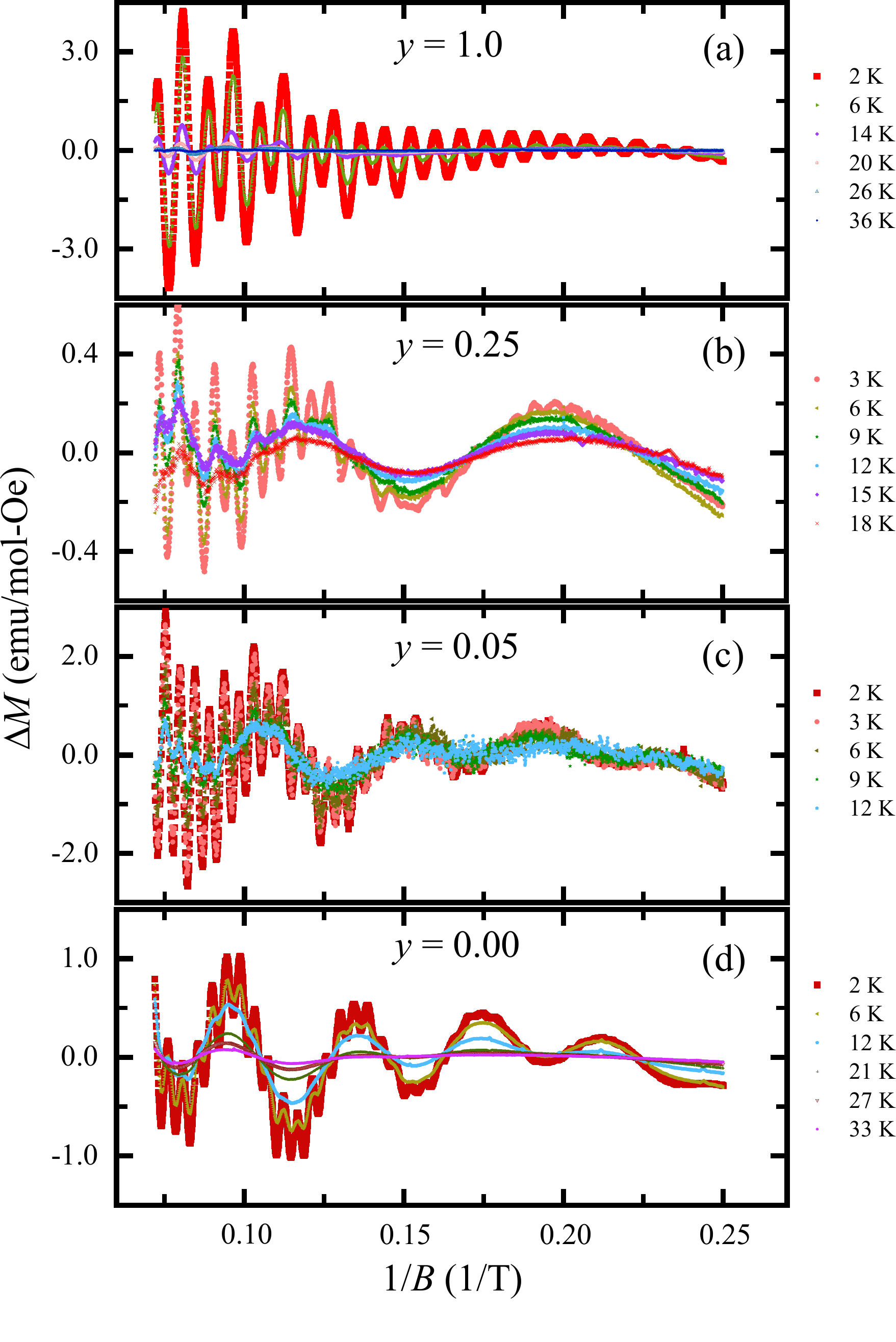}
    \caption{
    Oscillatory component of magnetization $\Delta M$ of $y$ = 1.00 (a), $y$ = 0.25 (b), $y$ = 0.05 (c) and $y$ = 0.00 (d) samples of \lit \hspace{1 pt} obtained after background subtraction from the data in Fig. \ref{fig:VSM} (b). The evolution of the dHvA oscillations with increasing Lu concentration indicate an increasingly complicated Fermiology in the vicinity of the CDW transition.
    }
    \label{fig:DMvInvB}
\end{figure}

Further investigation of SdH oscillations were not viable owing to limited sample dimension along the [001] direction. So instead, the evolution of the Fermiology with y across \lit \hspace{1 pt} can be tracked via change in magnetization ($\Delta M$) vs. $1/B$ in Fig. \ref{fig:DMvInvB} for $y$ = 1.00 (a), $y$ = 0.25 (b), $y$ = 0.05 (c), and $y$ = 0.00 (d), of which a similar polynomial background subtraction was applied to the $M$ vs. $B$ data in Fig. \ref{fig:VSM} (b) as was done for the $R$ vs. $B$. Although only the $\alpha$ and $\beta$ frequencies are visually apparent in the $y$ = 1.00 data, the $y$ = 0.25 results clearly have a 3rd very low frequency that is consistent with the very faint $\gamma$ feature identified in the FFT of $\Delta R$ in Fig. \ref{fig:DR_FFT_EffectiveMass} (a). All three frequencies increase with increasing Lu fraction, but the relative amplitude between the three components change in relation to each other with the $\alpha$ orbit being dominant at the Y-rich end and $\gamma$ being the dominant one at the Lu-rich end, which indicates a complicated interplay between scattering, density of states, and effective mass across multiple Fermi surfaces.

 \begin{figure}[]
    \centering
    \includegraphics[scale=0.23]{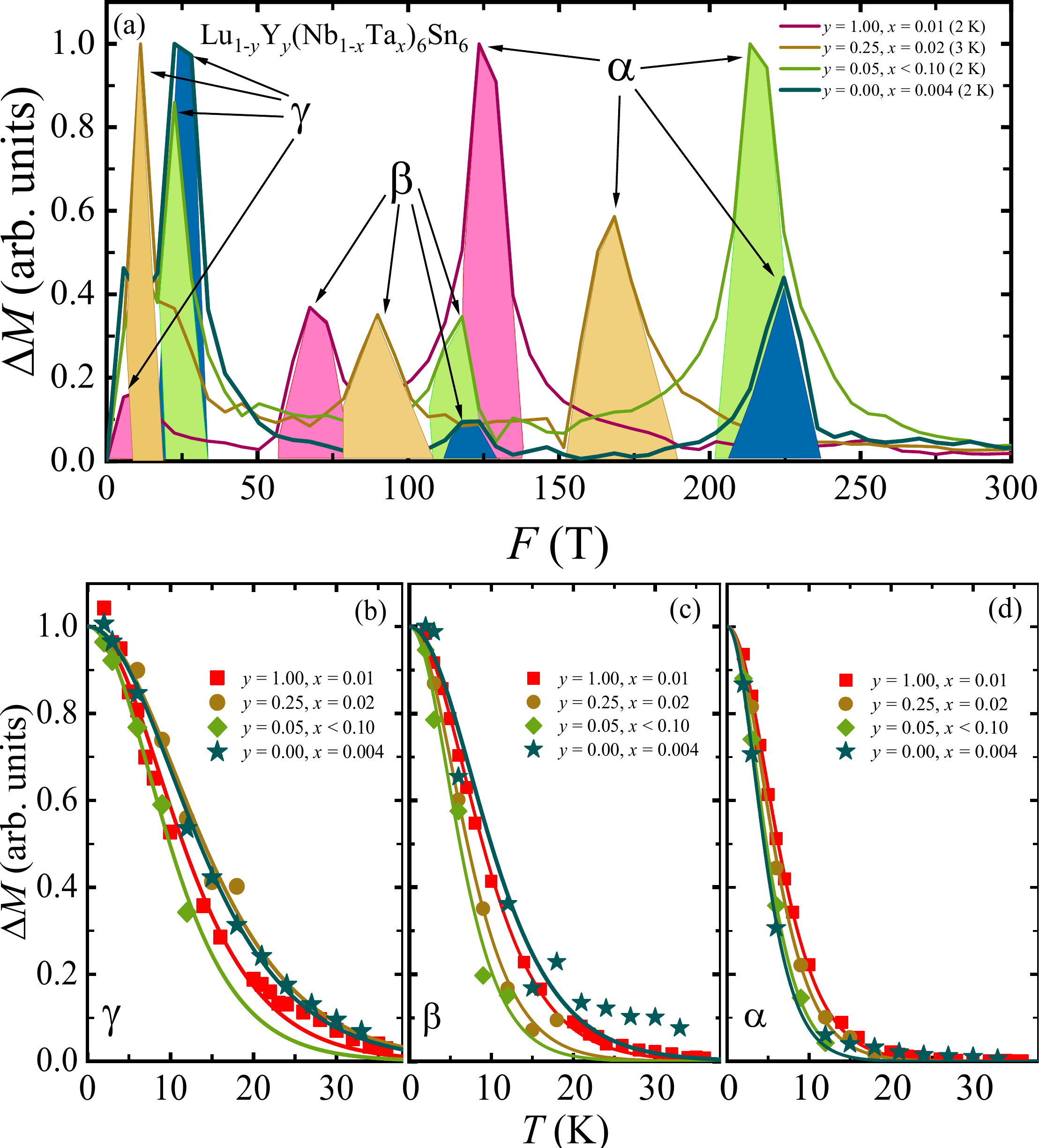}
    \caption{
    (a) Collective Fourier transform spectra of oscillatory component of magnetization $\Delta M$ data plotted in Fig. \ref{fig:DMvInvB}. All three frequencies, $\alpha$, $\beta$, and $\gamma$, shift to higher frequencies as Lu content increases. The relative amplitude of the middle frequency $\beta$ remains minor, but there is a clear change from the high frequency $\alpha$ peak being dominant at the Y end while the low frequency $\gamma$ component becomes the greatest contribution at the Lu end, suggesting a density of states transfer coinciding with the CDW order. The retroactively normalized $\Delta M$ vs $T$ of the $\gamma$ (b), $\beta$ (c) and $\alpha$ (d) FFT components across the varying Y alloying is used to extract the effective mass by fitting the data as described in the text.}
    \label{fig:DM_FFT_EffectiveMass}
\end{figure}

The FFT of the $\Delta M$ vs $1/B$ at 2 K are shown in Fig. \ref{fig:DM_FFT_EffectiveMass} (a) and all data are normalized to the magnitude of the highest peak with the range being from 14 T (0.0714 T$^{-1}$) and 4 T (0.25 T$^{-1}$). The shaded regions are guides to the eye and confirm that the aforementioned frequencies increase with Lu fraction and the gradual weight shift among the components. Investigation of their temperature dependence are shown for the $\gamma$ (b), $\beta$ (c), and $\alpha$ (d) frequencies with all data retroactively normalized to 1 at 0 K from fitting eqn. \ref{eqn:TemperatureOnly} with the solid curves as the fits. There is minimal systematic variations apparent among the $\gamma$ and $\beta$ components, although minor changes in the decay with temperature are apparent. The outlier is the $\alpha$ orbit with the largest effective mass in which there is a slight, but systematic increase in decay with $T$ that indicates the effective mass increases with Lu content.

 \begin{figure}[]
    \centering
    \includegraphics[scale=0.32]{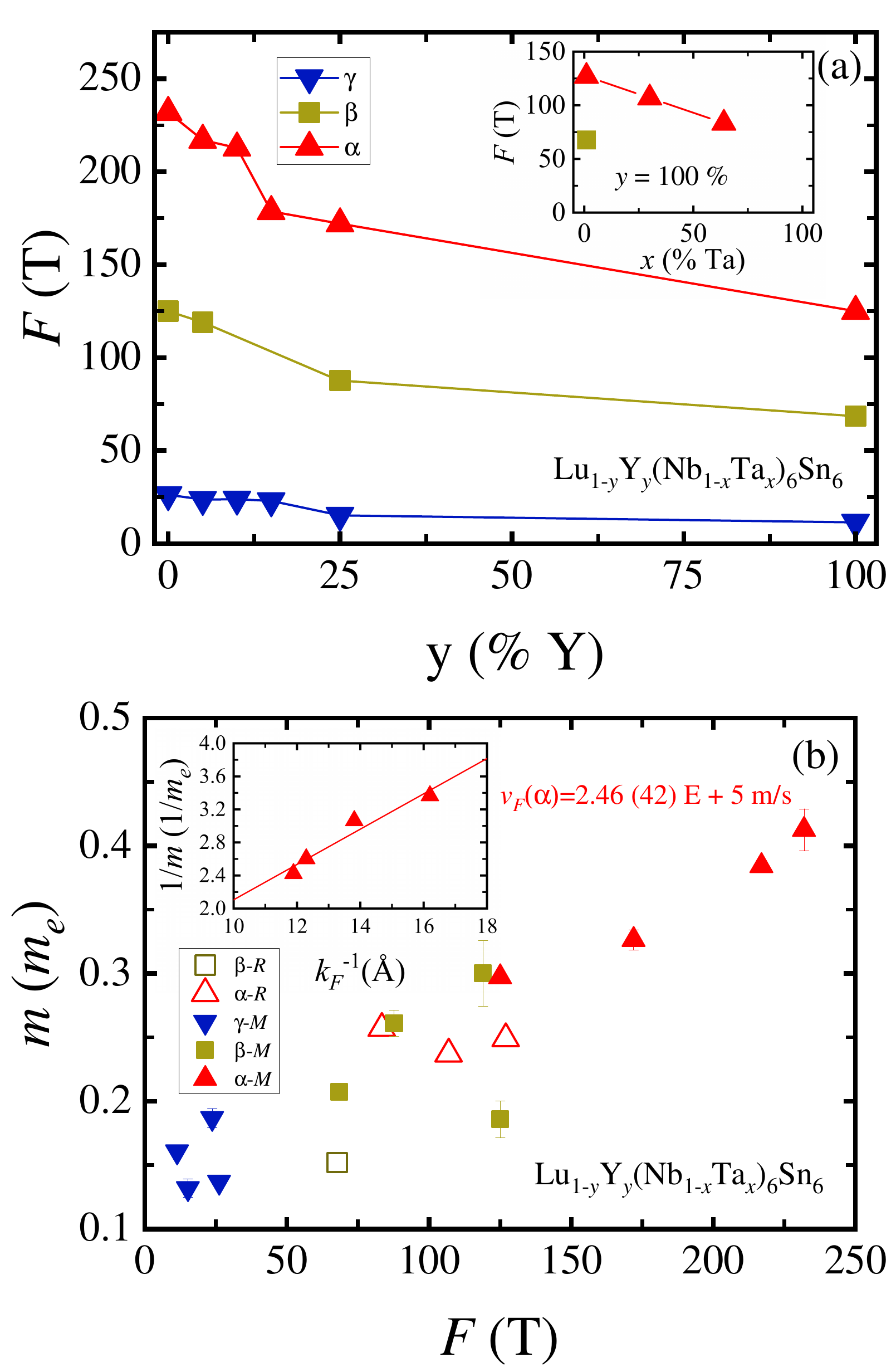}
    \caption{(a) Phase diagram of oscillation frequencies in \lit\ as a function of Y concentration.
    Frequencies monotonically increase with decreasing Y\% until $\sim$15\%, where a more abrupt increase coincides with the onset of CDW order. The inset shows frequency dependence on Ta\% ($x$) of \yit\ for comparison. 
    (b) Evolution of effective mass extracted from quantum oscillations data as a function of orbit frequency $F$ for available $x$ and $y$ alloying,  demonstrating an approximately linear correlation. The inset shows the relation between inverse quantities ($1/m$ vs  inverse Fermi wave-vector $k^{-1}_F$) for the $\alpha$ orbit resolved through dHvA alongside a linear fit to a Dirac dispersion as described in the text that allows extraction of a Fermi velocity ($v_F$).}
    \label{fig:FvY_MvF}
\end{figure}

The important takeaways from analysis of the SdH and dHvA oscillations across available \lit \hspace{1 pt} samples are in Fig. \ref{fig:FvY_MvF}, which show the relations between $F$ and y (a), and $m$ vs. $F$ (b). As the FFT process introduces broadening owing to finite signal processing issues, we determine $F$ by appropriate band-pass filtering the $M$ vs. $1/B$ data and manually counting peak to peak distances in 1/T for each orbit of the dHvA oscillations.  The change in $F$ appears minimal from $y$ = 1.00 to $y$ = 0.25, although we lack results from $y$ = 0.25 to $y$ = 1.00, there is not expected to be any phase transitions that could induce Fermi surface changes in this region, which is supported by the lack of any noticeable aberrations in the aforementioned $\rho$ and $M$ data. Between $y$ = 0.25 to $y$ = 0.00 is where the most apparent changes to the Fermi surfaces occur with a minor bump in $F_{\gamma}$ from $y$ = 0.25 to $y$ = 0.15 until it levels off to $\sim$ 25 T. Any abrupt behavior of $F_{\beta}$ is a bit harder to investigate owing to its small amplitude and lacking higher field data for the $y$ = 0.15 and 0.10 samples, but nonetheless its increase monotonically tracks with the other orbits. The most sudden increase occurs in $F_{\alpha}$ in which it jumps from $\sim$ 175 T to above 200 T from a minor change in $y$ from 0.15 to 0.10 and may be due to influence from the CDW order upon which it continues to increase until $y$ = 0.00. The inset of Fig. \ref{fig:FvY_MvF} (a) shows the frequencies from SdH oscillations as a function of $x$ with Ta substitution for the $y$ = 1.00 end of the data and demonstrates an apparently linear in $x$ decrease of $F_{\alpha}$ with no additional deviations, which is consistent with the lack of any new phases emerging with Ta alloying.

The correlation between $m$ and $F$ is less straightforward. There is a clear trend of $m_{\gamma} < m_{\beta} < m_{\alpha}$ with the occasional outlier, but within each orbit things can vary. The $\gamma$ orbit demonstrates no obvious correlation between the two, but was also observable in low field SdH measurements of ScV$_6$Sn$_6$ with a similar effective mass \cite{DeStefan2023QuantumMat_Sc166_LowFieldQO}. The $\beta$ orbit shows a positive correlation at first, but the highest frequency data point show an abrupt drop in $m$, although as this corresponds to the $y$ = 0.00, $x$ = 0.004 sample, it may simply by an inaccuracy related to the reduced amplitude of the $\beta$ orbit in that sample. The $\alpha$ orbit shows the widest extent spanned in $F$, and indicates that while $m$ is mostly constant below 125 T, it  increases with larger Fermi surface between 125 T and 230 T, which is fully incompatible with a parabolic band structure in which $m$ is constant with $F$. If this is indeed from a linear dispersing Dirac point, than we take the approach used in treating RhSb$_3$ \cite{Wang2023_PRM_RhSb3} and linear fit $1/m$ vs inverse Fermi wave number ($k^{-1}_F$) as demonstrated in the inset, with $k^{-1}_F$ being determined from the Onsager rule and the spherical Fermi surface approximation. This approach gives a relation $1/m \approx v_F/(\hbar k_F)$ \cite{Castro2009RevModPhys.81.109GRaphene} with the Fermi velocity as the slope in appropriate units, of which converts to the listed value of 2.46 E + 5 m/s.  It is possible that deviations from apparent linearity in the fit could be due to the simplistic way in which effective mass was obtained, as fitting multiple frequency versions of Eqn. \ref{eqn:FullLK} is non-trivial, but the other is that the presence of CDW order induces a minute change to the relativistic Dirac dispersion. In the opposite direction, as the $\alpha$ Fermi surface shrinks below 100 T the effective mass trends towards a constant, which suggests more parabolic band structure and conventional behavior.


 \begin{figure}[]
    \centering
    \includegraphics[scale=0.29]{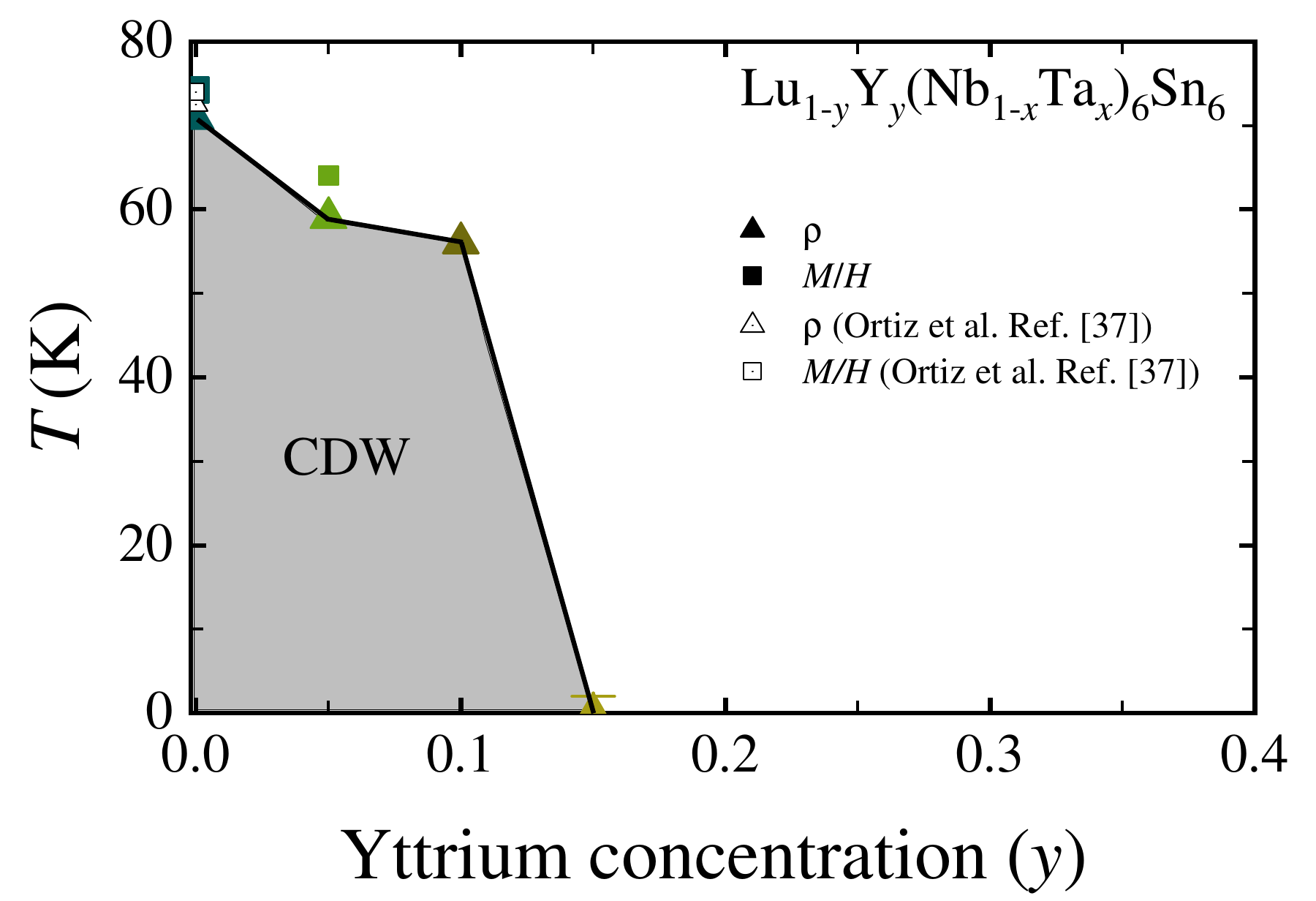}
    \caption{(a) Evolution of charge density wave (CDW) onset transition temperature with Y concentration in \lit. As changes in $x$ have minimal affect, we only show samples with the smallest Ta amount where applicable. The effect of Y substitution for Lu has a moderate suppression on the CDW order, until its signatures in $\rho$ and $M/H$ abruptly broadened and disappear between $y$ = 0.10 and $y$ = 0.15, which suggests the phase transition is of a discontinuous nature. Data points determined from highest temperature kinks or deflection in $\rho$ $(T)$ and $M$ $(T)$ and indicate the onset of the CDW order. 
    }
    \label{fig:PHASE}
\end{figure}

The CDW phase diagram constructed from the $\rho$ and $M$ data is shown in Fig. \ref{fig:PHASE}, of which the color scheme of the data points matches those taken from Figs. \ref{fig:RhoAndMRFromYtoLu} and \ref{fig:VSM}. As the Ta only minutely suppresses the CDW order, we only show the samples with the lowest Ta amounts. This is in direct contrast to Sc(V$_{1-x}$Cr$_x$)$_6$Sn$_6$ in which a few percent Cr eliminates the CDW order \cite{Lee2024Nature_Sc166_ARPES_CDW_surpressWithChromium}, which indicates that the underlying mechanism of the CDW could be different between LuNb$_6$Sn$_6$ and ScV$_6$Sn$_6$, the only two materials of this 166 kagome family with charge order, but we note that Cr has one additional valence electron relative to V, whereas Nb and Ta are isovalent, and the CDW suppression in this case could simply be related to changing the unit cell size. Our LuNb$_6$Sn$_6$ CDW transition temperature is comparable with the equivalent signatures from Ref. \cite{Ortiz2025ACS_OriginalSynthesisLuNb6Sn6}. Although the exact details of the CDW order is an ongoing investigation, it is quite apparent that the transition is not smoothly suppressed to zero temperature. This suggests that the transition is only broadened and/or becomes first order before any features of quantum criticality can become realized.

\section{Discussion}

All in all, the \lit \hspace{1 pt} material provides a clean platform in which to study the interplay between Dirac band dispersion, disorder, and CDW order.  The abrupt increase in Fermi surface size inferred from the Dirac $\alpha$ orbit frequency in the vicinity of the CDW transition guaranties as such, although opposite in behavior to CsV$_3$Sb$_5$ quantum oscillations under pressure in which it is observed that the small Fermi surface volumes increase as the CDW is suppressed with pressure alongside fluctuation enhanced effective masses at the critical suppression pressure \cite{Zang2024PNAS_Cs135}. We also emphasize that this can not be an issue of adding holes nor electrons as Y and Lu are isoelectronic to each other with the same number of valence electrons and stable oxidation states (same for Nb and Ta). Instead, the change in Dirac Fermi surface must be tuned by competing interactions, which has been investigated experimentally \cite{Lee2021AdvMat_DiracSemiMetal_CDW_GdSbTe} and theoretically \cite{ChenPRL2019_MonteCarlo_DiracAndCDW_QCP} in other Dirac systems. This may also be why the $\alpha$ Fermi surface becomes less prominent compared to the $\gamma$ surface at the Lu end; a gradual shift of density of states from one to the other. This is especially important as the effective mass of the $\gamma$ orbit shows no increase nor decrease with Fermi surface size, which suggests that it has a dispersion closer to quadratic. It is an open issue as to the implications of the LuNb$_6$Sn$_6$ CDW order \cite{Yang2025PRB_Lu166_ARPES_Xray_FrustraitedCDW, Lou2025_arxiv_Lu166_OrbitalSelectiveStuff_STM_ARPES} inducing transfer from one band to another in the presence of non-trivial topology. The CDW order in LuNb$_6$Sn$_6$ seems to be driven by a subtle rattling mode of the rare-earth and Sn chains of the system, such as seen in ScV$_6$Sn$_6$ \cite{Meier2023_ScV6Sn6_CDW_rattling}, which is consistent with why our Ta has minimal effect on such an order as the Nb/Ta ions are not involved with such a mode, but are still easily able to modify the band structure. 

Even without an ordered ground state, quantum oscillations in Kagome lattice systems have a long history. Both isostructural LuV$_6$Sn$_6$ \cite{Philips2025PRB_LuV6Sn6_HighFieldTorqueAndQuantumOscillations} and YV$_6$Sn$_6$ \cite{Shtefiienk2025APLquantum_YV6Sn6_45TQO} exhibit quantum oscillation frequencies consistent with our $\alpha$ and $\gamma$ orbits at the Lu end of \lit, although require high magnetic field facilities in order to observe. Extensive work into the Chern number of TbMn$_6$Sn$_6$ was viable via quantum oscillations \cite{Yin2020Nature_TbMn6Sn6_ChernAndQuantumOscillations} and anomalous Hall effect \cite{Xu2022Nature_TbMn6Sn6AHE}. Changes in quantum oscillation relative amplitudes, but not frequencies, have been observed between CsTi$_3$Bi$_5$ and RbTi$_3$Bi$_5$ \cite{Rehfuss2024PRM_Cs135_Rb135} despite Cs and Rb being isovalent, which is at odds with our Lu to Y and Nb to Ta comparisons and point to the change in frequencies being more than simple changes of the Fermi energy.  Of course, exact Fermiology is not completely determined by the crystal structure and electron counting per atom in practice, with well know examples of electron phonon interactions causing the details of the band structure to be renormalized \cite{Giovannini2020PhysRevLett_TimeResolvedArpes, Myasnikova_2018}, which could indicate that the CDW instability has a similar effect on the Dirac Fermi surface in \lit. A simple toy model argument can be constructed by observing that a gaped Dirac point energy dispersion has a form of $E = \sqrt{\hbar^2 v^2_F k^2 + \Delta ^2}$ \cite{DRAGOMAN2024108818} with the gap, $\Delta$, usually being non-zero due to spin-orbit coupling \cite{Freitas2026MatSci_DiracGap, 1pwr-lzz5, YAKOVKIN20171}. A plausible reason for why the substitution of Ta causes a decrease in Fermi surface size in \yit \hspace{1 pt} is simply that 5$d$ Ta has stronger spin-orbit coupling than 4$d$ Nb, which decreases the density of states via enchantment of $\Delta$. Such a spin-orbit effect is clear in previous density functional theory calculations comparing YV$_6$Sn$_6$, YNb$_6$Sn$_6$, and YTa$_6$Sn$_6$ (if $x$ = 1.00 can exist) in which the K-point Dirac gap is enhanced from 27.2 meV for Nb to 112 meV for Ta \cite{LanTin2023PRB_TheoryYT6Sn6}. The enhancement of Fermi surface size with Lu substitution is less clear and suggests we are enhancing the band dispersion via changing the slopes of the bands in this case and is at odds with the RhSb$_3$ \cite{Wang2023_PRM_RhSb3} scenario in which defects and impurities  could act as a true carrier dopent source to shift the chemical potential. Regardless of the microscopic details, tracking effective mass with Fermi surface size was used to verify the existence of  relativistic Dirac Fermions in the seminal work on graphene \cite{Novoslev2005Nature_Graphen}, although in our case of \lit \hspace{1 pt} we are dealing with a more 3D system which has a non-zero $\Delta$, which provides a platform for studying Dirac Fermions at the boundary between non-relativistic and relativistic regimes.

\section{Conclusion}

In conclusion, we have detailed our growth and characterization of kagome lattice \lit \hspace{1 pt} double-alloyed single crystals. The electrical transport and magnetization results demonstrated that they are clean with semi-metallic character and are tunable through a first-order CDW instability. Quantum oscillations show a Fermi surface size dependent effective mass indicative of  the existence of a non-parabolic Dirac band structure. Investigation of the quantum oscillations yields a relation between the onset of CDW order and an abrupt increase in Fermi surface volume that provides a platform for exploration of CDW order interacting with relativistic Dirac Fermions as well as how those Dirac Fermions subtly transform to non-relativistic massive particles at smaller Fermi surface volumes.

\begin{acknowledgments}
Research at the University of Maryland was supported by
the Gordon and Betty Moore Foundation’s EPiQS Initiative Grant No. GBMF9071,
the U.S. National Science Foundation Grant Nos. DMR2303090 and DMR2523217, 
the NIST Center for Neutron Research, and the Maryland Quantum Materials Center. 
\end{acknowledgments}

%

\end{document}